\documentclass[aps,prl,twocolumn,preprintnumbers,amsmath,amssymb,
superscriptaddress,showpacs,floatfix]{revtex4}

\usepackage{graphicx}
\usepackage{dcolumn}
\usepackage{bm}
\usepackage{epsfig}
\usepackage{graphics}
\usepackage{latexsym}

\def\beq{\begin{equation}}
\def\eeq{\end{equation}}
\def\bea{\begin{eqnarray}}
\def\eea{\end{eqnarray}}
\newcommand{\beqs}{\begin{subequations}}
\newcommand{\eeqs}{\end{subequations}}

\newcommand{\cref}[1]{Ref.~\cite{#1}}

\newcommand{\vev}[1]{\left<#1\right>}

\newcommand{\Tr}{\mathsf{Tr}}

\newcommand{\hh}{{\ensuremath{I{\kern-2.6pt h}}}}
\newcommand{\bhh}{{\ensuremath{\bar{I{\kern-2.6pt h}}}}}

\begin{document}


\title{Electroweak monopoles and magnetic dumbbells in grand unified 
theories}

\author{G. Lazarides}
\email{glazarid@auth.gr} 
\affiliation{School of Electrical and
Computer Engineering, Faculty of Engineering, Aristotle University
of Thessaloniki, Thessaloniki 54124, Greece}
\author{Q. Shafi}
\email{shafi@bartol.udel.edu}
\affiliation{Bartol Research Institute, Department of Physics and 
Astronomy, University of Delaware, Newark, Delaware 19716, USA}

\date{\today}

\begin{abstract}
We use the $SU(5)$ model to show the presence in grand 
unified theories of an electroweak monopole and a magnetic 
dumbbell (``meson'') made up of a 
monopole-antimonopole pair connected by a $Z$-magnetic flux tube.
The monopole is associated with the spontaneous breaking of 
the weak $SU(2)_L$ gauge symmetry by the induced vacuum 
expectation value of a heavy scalar $SU(2)_L$ triplet with 
zero weak hypercharge contained in the adjoint Higgs 24-plet. 
This monopole carries a Coulomb magnetic charge of $(3/4)
(2\pi/e)$ as well as $Z$-magnetic charge, where $2\pi/e$ 
denotes the unit Dirac magnetic charge. Its total magnetic 
charge is $\sqrt{3/8}(4\pi/e)$, which is in agreement with 
the Dirac quantization condition. The monopole weighs about 
700 GeV, but because of the attached $Z$-magnetic tube it 
exists, together with the antimonopole, in a magnetic 
dumbbell configuration whose mass is expected to lie in the 
TeV range. The presence of these topological structures in 
$SU(5)$ and $SO(10)$ and in their supersymmetric extensions 
provides an exciting new avenue for testing these theories 
in high-energy colliders. 
\end{abstract}

\maketitle

Grand unified theories based on the gauge groups $SU(5)$ 
\cite{GeorgiGlashow}, $SO(10)$ \cite{minkowski}, and $E_6$ 
\cite{ramond} all predict the existence of a superheavy and 
topologically stable magnetic monopole, which carries a 
single quantum ($2\pi/e$) of Dirac magnetic charge \cite{dirac}
as well as color magnetic charge \cite{dokos,daniel}. In $SU(5)$ 
this is the lightest monopole, but $SO(10)$ and $E_6$ may also give 
rise to topologically stable intermediate-scale monopoles depending 
on their symmetry-breaking patterns. For instance, $SO(10)$ breaking 
via $SU(4)_c\times SU(2)_L\times SU(2)_R$ \cite{PatiSalam} yields 
an intermediate-scale topologically stable monopole carrying two 
units of the Dirac charge as well as color magnetic charge 
\cite{magg}. Depending on its 
symmetry-breaking scale, the trinification symmetry group $SU(3)_c
\times SU(3)_L\times SU(3)_R$ can yield a topologically stable triply 
charged monopole with mass in the TeV region, thus making it 
potentially accessible at high-energy colliders \cite{TripMon}.

In this paper, we display the presence of another class of topological 
structures that can appear in grand unified theories. These structures
are not topologically stable, and, as we show using the $SU(5)$ example, 
they are magnetic dumbbells (or ``mesons'') made up of electroweak 
monopole-antimonopole pairs connected by magnetic $Z$-flux tubes. The 
magnetic dumbbell, with mass estimated to lie in the TeV range, is 
closely related, it appears, to objects of essentially the same name 
discussed some time ago by Nambu \cite{Nambu} in the standard 
electroweak model. However, there is an important difference between 
these two cases, which is related to the fact that the electroweak 
monopole in $SU(5)$ is associated with the induced vacuum expectation 
value (VEV) of a heavy $SU(2)_L$ triplet field with zero weak 
hypercharge $Y$ that resides in the adjoint Higgs 24-plet. There is, of 
course, no corresponding elementary $SU(2)_L$ triplet scalar in the 
standard $SU(2)_L\times U(1)_Y$ model. The electroweak monopole in
$SU(5)$ carries a Dirac charge of $(3/4)(2\pi/e)$ as well as 
$Z$-magnetic charge, and it is expected to exist in a confined state 
together with its antimonopole. (Monopoles carrying Coulomb and 
$Z$-flux with the aim of confining primordial monopoles were previously 
discussed in Ref.~\cite{Ztube}. For two basic papers on electroweak 
strings, see Ref.~\cite{tanmay}.) 

We should emphasize that our results also apply to extensions of 
the minimal $SU(5)$ model that incorporate additional matter and/or 
scalar fields to implement gauge coupling unification and improve 
fermion mass relations, and are compatible with the proton lifetime 
limits -- see, for example, Ref.~\cite{rehman}.

It is important to clarify that the $SU(5)$ model does not predict 
topologically stable electroweak monopoles or strings associated 
with the electroweak breaking. To this end, we concentrate on the 
electroweak sector of the model, where the 
generator of $U(1)_Y$ is taken as $2Y$ so that it has integral 
eigenvalues and, thus, periodicity $2\pi$. We note that inside 
$SU(5)$, the electroweak gauge symmetry is $G=SU(2)_L\times 
U(1)_Y/Z_2$, where $Z_2$ is generated by the element $(-1,-1)=
(e^{i\pi T^3_L},e^{i\pi 2Y})$ of $SU(2)_L\times U(1)_Y$ with 
$T^3_L={\rm diag}(1,-1)$. Indeed, this element acts as the 
identity element on the $SU(2)_L$ doublet 
with $Y=-1/2$ and, consequently, on all the representations of 
the model including the real triplet with $Y=0$, as well as any 
other  complex triplets with integral hypercharge. The second 
homotopy group of the vacuum manifold is
\beq
\pi_2\left(\frac{G}{U(1)_{em}}\right)=
\pi_1\left(U(1)_{em}\right)_G, 
\label{homopi2}
\eeq      
where the right-hand side consists of all the homotopically 
nontrivial loops in the electromagnetic gauge group 
$U(1)_{em}$ which are trivial in $G$. The smallest nontrivial 
loop in $U(1)_{em}$ corresponds to a $2\pi$ rotation along the 
electric charge generator $Q=(T^3_L+2Y)/2$, which is equivalent 
to a $\pi$ rotation along $T^3_L$ accompanied by a $\pi$ 
rotation along $2Y$ interpolating between $(1,1)$ and 
$(-1,-1)$ in $G$. This is the smallest closed loop in $G$ and 
is homotopically nontrivial. The fundamental (first homotopy) 
group in the right-hand side of Eq.~(\ref{homopi2}) is 
therefore trivial, and there are no topologically stable 
electroweak monopoles. Moreover, the fundamental group of the 
vacuum manifold
\beq
\pi_1\left(\frac{G}{U(1)_{em}}\right)=
\pi_0\left(U(1)_{em}\right)_G 
\label{homooi1}
\eeq     
is also trivial, since both $G$ and $U(1)_{em}$ are connected,
and no stable strings appear either.    

In the electroweak model, we introduce a real Higgs triplet field 
$T=T_i\sigma_i/2$ with a vanishing hypercharge $Y$, where $\sigma_i$ 
($i=1,2,3$) are the three Pauli matrices. This triplet, which 
resides in the $SU(5)$ adjoint Higgs 24-plet, couples to the 
electroweak doublet $H$, as displayed in the following additional
contribution to the potential energy density:
\beq
V_T=\frac{1}{2}M_T^2\left(T_i-\frac{\lambda_T}{M_T}H^{\dagger}
\sigma_i H\right)^2.
\label{pot}
\eeq 
Here, $M_T\gg M_Z$ is the triplet mass, which may be as large as the 
grand unification scale, and $\lambda_T$ is a dimensionless coupling 
constant of the order of unity or less. The cross term in
the above expression originates from the Higgs couplings $5^\dagger
\times 24\times 5$ and $5^\dagger\times 24^2\times 5$ \cite{ellis} 
between the $SU(5)$ Higgs 24-plet and the Higgs $5$-plet which 
contains the electroweak doublet $H$. It yields a $T$ VEV suppressed
relative to the VEV $\vev{H}$ of $H$ by a factor $\vev{H}/M_T$. In 
Eq.~(\ref{pot}) we have left out quartic scalar couplings such as 
$(\Tr(T^2))^2$ and $(H^\dagger H)\Tr(T^2)$ which do not play an 
essential role here.

The spontaneous breaking of the electroweak symmetry is achieved, 
as usual, via the potential
\beq
V_H=\frac{\lambda}{4}\left(H^\dagger H-\frac{v_D^2}{2}\right)^2,
\label{eq:VH}
\eeq
which yields the following VEV for the electroweak doublet $H$:   
\beq
\vev{H}=\begin{pmatrix} \frac{v_D}{\sqrt{2}} \\& \\ 0
\end{pmatrix},
\eeq
where $v_D\simeq 246~{\rm GeV}$. From Eq.~(\ref{pot}), this 
induces the triplet VEV given by 
\beq
\vev{T_3}=\frac{\lambda_T v_D^2}{2M_T}\equiv v_T,
\label{tripvev}
\eeq
which breaks $SU(2)_L$ to its $U(1)_L$ subgroup with generator 
$T^3_L={\rm diag}(1,-1)$. 

Ignoring for the moment the electroweak symmetry breaking by 
$\vev{H}$, the breaking of $SU(2)_L$ by the 
Higgs triplet $T$ yields a 't Hooft-Polyakov-type monopole 
\cite{hooft} with magnetic flux corresponding to a $2\pi$ 
rotation around $T^3_L$ or, equivalently, a $4\pi$ rotation 
around the customarily normalized generator $T^3_L/2$. 
Reintroducing $\vev{H}$, this monopole ceases 
to be topologically stable and becomes attached to a magnetic 
flux tube. Indeed, the electroweak symmetry breaking leaves
unbroken the electric charge generator $Q= T^3_L/2 + Y$, 
where the weak hypercharge operator is given by 
$Y={\rm diag}(-1/3,-1/3,-1/3, 1/2, 1/2)$ in $SU(5)$. The 
corresponding orthogonal broken generator is 
$\mathcal{B}=T^3_L/2-3Y/5$. 

At this stage, it is convenient to consider $5\mathcal{B}$, 
which has 
the smallest possible integer elements and, thus, periodicity 
$2\pi$. A rotation by $2\pi/4$ along $5\mathcal{B}$ leaves 
invariant the VEV of $H$, and, therefore, the associated tube 
carries $Z$-magnetic flux corresponding to a $2\pi$ rotation 
around $5\mathcal{B}/4$. From the relation $T^3_L=3Q/4+
5\mathcal{B}/4$, we see that the monopole with one unit of 
flux along $T^3_L$ (i.e. corresponding to a $2\pi$ rotation 
around this generator) is attached to a $Z$-flux tube with 
one unit of flux along $5\mathcal{B}/4$ and also has Coulomb 
flux $3Q/4$. The magnetic 
charge corresponding to the Coulomb flux of the monopole is 
$g_M=(3/4)(2\pi/e)$. A monopole and an antimonopole are 
expected to pair up and form a dumbbell connected by this 
flux tube. 


A few remarks about the $Z$-magnetic flux emerging from the 
electroweak monopole  are in order here. The normalized 
generator orthogonal to $Q$ is $\sqrt{5/8}(T^L_3/2-3Y/5)$. 
As we have shown, the $Z$-magnetic flux in the tube corresponds 
to a $2\pi$ rotation around $(5/4)(T^L_3/2-3Y/5)$. Consequently, 
this flux is $(2\pi/g)\sqrt{5/2}$, where $g$ is the $SU(5)$ 
gauge coupling which, in the $SU(5)$ limit, coincides with the 
$SU(2)_L$ gauge coupling $g=e/\sqrt{3/8}$. The magnetic flux 
along the $Z$ tube therefore takes the form $(4\pi/e)\sqrt{3/8}
\sqrt{5/8}$. For completeness, we should note that the expressions 
above for the Coulomb and $Z$-magnetic fluxes of the electroweak 
monopole coincide, respectively, with the values 
$4\pi\sin^2\theta_W/e$ and $4\pi\sin\theta_W\cos\theta_W/e$ 
found by Nambu \cite{Nambu} by recalling the $SU(5)$ prediction 
$\sin^2\theta_W=3/8$, where $\theta_W$ is the electroweak angle. 
Combining appropriately these two fluxes, one obtains the total 
$SU(2)_L$ magnetic charge $4\pi/g$ of the electroweak monopole,
in full agreement with the results found by Nambu \cite{Nambu} 
and Vachaspati \cite{vacha}.

To reconfirm that the electroweak monopole accompanied by a 
$Z$-magnetic flux tube provides a consistent description, let us 
take the left-handed neutrino with zero electric charge around 
this tube. If the neutrino is covariantly transported, 
its wave function acquires an Aharonov-Bohm phase given by 
$\exp(i Q_Z^\nu \Phi_Z)$, where $Q_Z^\nu$ denotes the neutrino Z 
charge and $\Phi_Z$ denotes the $Z$-magnetic flux in the tube. 
Substituting 
$Q_Z^\nu=(e/\sin\theta_W\cos\theta_W)(T^3_L/2-Q\sin^2\theta_W)$ 
with $T^3_L=+1$, $Q=0$, and requiring the wave function to be 
single valued shows that the $Z$-magnetic flux $\Phi_Z$ is 
quantized in units of $(4\pi/e)\sin\theta_W\cos\theta_W$, 
in agreement with the discussion above. A related calculation for the 
charged leptons and quarks also takes into account the ordinary 
Coulomb magnetic flux which is carried by the electroweak monopole. 
As an example, for the left-handed d 
quark kept within its confinement radius $\sim\Lambda_{QCD}^{-1}$ 
($\gg M_Z^{-1}$), the total phase acquired by its wave function, 
taking into account the $Z$ flux and Coulomb flux $\Phi_{em}$, is 
given by $(e/\sin\theta_W\cos\theta_W)(T^3_L/2-Q\sin^2\theta_W)\Phi_Z+
eQ\Phi_{em}$, with $T^3_L=-1$ and $Q=-1/3$. Substituting $\Phi_Z=
4\pi\sin\theta_W\cos\theta_W/e$ and requiring the wave function to be 
single valued yields $\Phi_{em}=(4\pi/e)\sin^2\theta_W$ as a solution. 
This is the desired value of the Coulomb magnetic flux which 
appropriately combined with the $Z$ flux of the monopole yields the 
$SU(2)_L$ magnetic flux of $(4\pi/e)\sin\theta_W$. Note that the wave 
function of an $SU(2)_L$ singlet quark or lepton acquires a zero 
overall phase under similar transport. This is consistent with the 
fact that the electroweak monopole is associated with the breaking 
of $SU(2)_L$ to $U(1)_L$.

In the above discussion based on $SU(2)_L\times U(1)_Y$, 
the value of the electroweak mixing angle $\theta_W$ is not predicted.
The situation in grand unified theories, however, is different, and the 
prediction $\sin^2\theta_W=3/8$ allows us to provide the magnitude of 
the $Z$-magnetic flux.

Finally, the quark confinement that we assumed above is 
not required for consistency in the presence of the electroweak 
monopole that also carries a $Z$-magnetic flux tube. However, we recall 
from the introduction that $SU(5)$ also predicts the existence of a 
topologically stable monopole that carries a Dirac charge of $2\pi/e$ 
as well as a screened color magnetic field.  Quark confinement in this 
case is required in order for the Dirac quantization condition to be 
satisfied beyond the screening radius $\sim \Lambda_{QCD}^{-1}$. 

Returning to Eqs.~(\ref{pot}) and (\ref{eq:VH}), we should note that, 
at tree level, the VEV of $H$ is not affected by the presence 
of the Higgs triplet $T$. Indeed, 
minimization of the combined potential $V=V_H+V_T$ is 
achieved at 
\beq
M_T^2\left(T_i-\frac{\lambda_T}{M_T}H^{\dagger}
\sigma_i H\right)=0,
\label{min1}
\eeq
and
\bea
&-&M_T^2\left(T_i-\frac{\lambda_T}{M_T}H^{\dagger}
\sigma_i H\right)\frac{\lambda_T}{M_T}H^{\dagger}\sigma_i
\nonumber \\
&+& \frac{\lambda}{2}\left(H^\dagger H-v_D^2\right)
H^\dagger=0. 
\label{min2}
\eea
In view of Eq.~(\ref{min1}), Eq.~(\ref{min2}) reduces to the
standard equation for the electroweak symmetry breaking:
\beq
\frac{\lambda}{2}\left(H^\dagger H-\frac{v_D^2}{2}\right)=0,
\eeq
and so the presence of the triplet $T_i$ does not affect the 
VEV of the doublet $H$. 

The $\rho$ parameter \cite{veltman} in our case is given by 
$\rho=1+4R^2$, with $R=v_T/v_D$ (see, e.g., Ref.~\cite{cruz}). 
From the 2$\sigma$ upper bound $\rho\lesssim 1.00077$ 
\cite{rho}, we find $v_T\lesssim 3.4~{\rm GeV}$. 
Equation (\ref{tripvev}) then implies that
\beq
M_T\gtrsim 4.4\left(\frac{\lambda_T}{0.5}\right){\rm TeV}.
\eeq
A triplet with mass in the TeV range could provide a new source 
for Higgs production at high-energy colliders. Moreover, the 
mixed quartic coupling involving $H$ and $T$ may be helpful in 
preventing the quartic Higgs coupling going to zero and, thereby, 
stabilize the electroweak vacuum. It would be interesting to 
explore these possibilities in more realistic models which, 
among other things, also implement gauge coupling unification.

With the ansatz $T_i=v_Tx_i/r$, where $x_i$ ($i=1,2,3$) are 
the spatial coordinates and $r$ is the radial distance, the 
potential in Eq.~(\ref{pot}) is minimized for
\beq
H^{\dagger}\sigma_i H=\frac{v_D^2}{2}\frac{x_i}{r}.
\label{hedgehog}
\eeq  
This is achieved by taking \cite{Nambu}
\beq
H=\frac{v_D}{\sqrt{2}}\begin{pmatrix} \cos\frac{\theta}{2} \\& 
\\ 
\sin\frac{\theta}{2}\,e^{i\varphi}
\end{pmatrix},
\label{higgs}
\eeq 
where $0\leq\theta<\pi$ and $0\leq\varphi<2\pi$ are the polar 
angles. It is important to note that
the formula in Eq.~(\ref{higgs}) has an ill-defined phase 
$\varphi$ on the negative $x_3$ axis where $\theta=\pi$. This 
reflects the fact that the monopole is accompanied by 
a string ($Z$-flux tube). (For a discussion of the 
stability of this string, see Ref.~\cite{stability} and papers 
listed therein. A careful analysis is required to assess the 
string stability in the presence of the scalar triplet. In the 
discussion above, we have seen that the $Z$-flux tube was required 
by the Dirac quantization condition. This leads us to conjecture 
that the string is susceptible to breaking through a 
monopole-antimonopole pair creation.)

To obtain a rough estimate of the monopole mass, following 
Ref.~\cite{Nambu}, we ignore for the moment the attached 
$Z$ tube and approximate the monopole by a sphere of 
radius $r$ within which the gauge fields, $H$, and $T$ are 
zero. (Being a heavy scalar field, we expect $T$ to approach 
its VEV inside an inner core of radius $M_T^{-1}$. The energy 
stored in this inner core can be ignored, which can be checked 
by examining the quartic and gradient terms for the triplet 
$T$.) Outside 
the sphere, all the Higgs fields lie in the vacuum, and we have 
a Coulomb magnetic field corresponding to the magnetic charge 
\cite{Nambu,vacha,ana,preskill}
\beq 
g_M=\frac{4\pi}{e}\sin^2\theta_W,
\label{gM}
\eeq
where $e$ is the absolute value of the electron charge. The 
energy of the monopole configuration is then
\beq
E_M=\frac{g_M^2}{8\pi r}+\frac{4\pi}{3}r^3V_0,
\eeq
where $V_0$, the potential energy density within the sphere, 
is given by 
\beq
V_0=\frac{\lambda v_D^4}{16}=\frac{m_H^2 v^2_D}{8}
\eeq
and $m_H=\sqrt{\lambda/2}\,v_D$ is the Higgs boson mass. 
The energy $E_M$ is minimized at 
\bea
r_{\rm min}&=&\left(\frac{g_M^2}{32\pi^2 V_0}\right)^
{1/4}=\sqrt{\frac{2}{e}}\sin\theta_W(m_Hv_D)^
{-1/2}\nonumber \\
&\simeq& 7\times 10^{-3}~{\rm GeV}^{-1},
\eea    
giving the monopole mass
\bea
m_M&\approx & \frac{2^{1/4} g_M^{3/2}V_0^{1/4}}{3\pi^{1/2}} =
\frac{2^{5/2}\pi \sin^3\theta_W(m_Hv_D)^{1/2}}{3e^{3/2}}
\nonumber \\
&\simeq & 688~{\rm GeV}. 
\eea 

One can calculate the $Z$-tube radius $\rho_{\rm str}$ and 
tension $\mu_{\rm str}$ following Ref.~\cite{Nambu}. We find
\beq
\rho_{\rm str}\simeq 1.86\times 10^{-2}~{\rm GeV}^{-1} 
\quad {\rm and} \quad \mu_{\rm str}\simeq 2.57\times 10^{5}
~{\rm GeV}^2.
\eeq
The string radius exceeds the monopole
radius by a factor of 2.5 or so. So it makes sense to consider a
string segment at least as long as its radius. The energy 
of the ``minimal'' string segment is about 4.8~TeV, which yields
a minimal dumbbell of 5.8~TeV, after including the potential 
energy from the Coulomb attraction between the 
monopole-antimonopole pair.

Nambu has argued \cite{Nambu} that a rotating relativistic 
dumbbell with energy $E$ and angular momentum $L$ may yield a 
Regge trajectory $L\sim \alpha_0^\prime E^2$, with 
$\alpha_0^\prime=1/(2\pi\mu_{\rm str})$. Using the relevant 
formulas in Ref.~\cite{Nambu}, we find that, for string lengths 
bigger than the minimal length, $L\gtrsim 35$ and 
$E\gtrsim 7.5~{\rm TeV}$. The dumbbell is expected to decay 
through the emission of photons, weak gauge bosons, hadrons and 
leptons with lifetime $(E/6.74)\times 10^{-5}~{\rm GeV}^{-1}
\gtrsim 7.33\times 10^{-27}~{\rm sec}$ \cite{Nambu}. The string 
can also decay by monopole pair creation with the decay rate per 
unit length 
\cite{stringdec}
\beq
\Gamma_{\rm str}=\frac{\mu_{\rm str}}{2\pi}e^{-\pi 
\frac{m_M^2}{\mu_{\rm str}}}\simeq 127~{\rm GeV}^2.
\eeq
The corresponding lifetime for a dumbbell with energy $E$ is 
$(E/5.13)\times 10^{-7}~{\rm GeV}^{-1}\gtrsim 9.64\times 
10^{-29}~{\rm sec}$, which is shorter than its radiative 
lifetime by about 2 orders of magnitude. (For a discussion 
on the evolution of dumbbells, see Ref.~\cite{urrestilla}.)

With the electroweak monopole mass estimated to be around 
700~GeV, it is plausible that highly unstable configurations, 
consisting of (overlapping) monopole-antimonopole pairs in a 
mass range of 1--2~TeV, may be produced in high-energy 
collisions.

Although we have discussed the presence of the electroweak 
monopole and TeV-scale magnetic dumbbells in minimal $SU(5)$, 
one could reasonably expect that similar structures also 
appear in larger grand unified theories. Indeed, based on our 
discussion, they seem to be a rather generic feature of such 
theories. Consider $SO(10)$, for instance, in which a nonzero 
VEV for the $SU(2)_L$ triplet $T$ will arise through the mixed 
quartic coupling $45^2\times 10\times 10$. Here, the 10-plet 
contains the standard Higgs doublet, and the 45-plet is the 
adjoint Higgs field. In a more elaborate $SO(10)\times 
U(1)_{PQ}$ model \cite{holman}, where $U(1) _{PQ}$ denotes the 
axion symmetry \cite{PecceiQuinn}, the trilinear mixed coupling 
$10\times 10\times 45$ will induce a nonzero VEV for the 
$SU(2)_L$ triplet in the 45-plet. The presence of TeV-scale 
magnetic dumbbells, made up of monopole-antimonopole pairs, thus 
appears to be a rather generic feature of grand unified theories.
Furthermore, the $SO(10)$ breaking to the standard model often 
proceeds through one or more intermediate steps. Suppose that 
the low-energy group, excluding QCD, is $SU(2)_L\times SU(2)_R 
\times U(1)_{B-L}$. In this case, we could proceed to break 
$SU(2)_R$ to $U(1)_R$ with an $SU(2)_R$ triplet scalar, which 
yields a monopole carrying $U(1)_R$ charge. With the next 
breaking of $U(1)_R\times U(1)_{B-L}$ to $U(1)_Y$, the monopole 
gets connected to a string (flux tube). We, therefore, expect the 
appearance of new dumbbells, and, if the monopole mass scale for 
this case is suitably large compared to the string scale, the 
flux tubes can be relatively stable and less likely to break 
via monopole-antimonopole pair creation. The dumbbell mass may 
lie in the TeV range depending on the symmetry-breaking scale 
of $SU(2)_L\times SU(2)_R\times U(1)_{B-L}$.

Let us mention that our discussion carries over to the 
supersymmetric extensions of the $SU(5)$ and $SO(10)$ models. 
The induced VEV for the scalar triplet $T$ arises in 
supersymmetric $SU(5)$, for instance, via the soft 
supersymmetry-breaking terms corresponding to the 
superpotential coupling 
$\bar{5}\times 24\times 5$, where $\bar{5}$ and 5 contain the 
two Higgs doublets. Consequently, this VEV is suppressed by 
an additional factor $m_0/M_T$, where $m_0$ denotes the soft
supersymmetry-breaking mass parameter. The electroweak monopole 
is therefore also expected to be present in these models. 
  
In summary, we have identified in $SU(5)$ the presence of an 
electroweak monopole that carries a Dirac magnetic charge of 
$(3/4)(2\pi/e)$ and a $Z$-magnetic flux. Under 
plausible assumptions, the monopole mass is estimated to be around 
700 GeV, and the associated $Z$-flux tube width and tension are of 
the order of $M_Z^{-1}$ and $30 M_Z^2$ respectively. 
The monopole-antimonopole pairs form dumbbells (mesons) discovered
some time ago by Nambu. A search for these extended structures at 
the LHC and its upgrades seems worthy of further consideration.
Finally, we have noted that analogous TeV-scale extended structures 
can also appear in larger gauge symmetries such $SO(10)$. For another 
discussion of the electroweak monopole with different conclusions, 
see Ref.~\cite{pqhung}.

\vspace{.3cm}


This work is supported by the Hellenic Foundation for Research 
and Innovation (H.F.R.I.) under the ``First Call for H.F.R.I. 
Research Projects to support Faculty Members and Researchers and 
the procurement of high-cost research equipment grant'' (Project 
No. 2251). Q.S. is supported in part by the Department of Energy 
Grant No. DE-SC-001380.

\def\ijmp#1#2#3{{Int. Jour. Mod. Phys.}
{\bf #1},~#3~(#2)}
\def\plb#1#2#3{{Phys. Lett. B }{\bf #1},~#3~(#2)}
\def\zpc#1#2#3{{Z. Phys. C }{\bf #1},~#3~(#2)}
\def\prl#1#2#3{{Phys. Rev. Lett.}
{\bf #1},~#3~(#2)}
\def\rmp#1#2#3{{Rev. Mod. Phys.}
{\bf #1},~#3~(#2)}
\def\prep#1#2#3{{Phys. Rep. }{\bf #1},~#3~(#2)}
\def\prd#1#2#3{{Phys. Rev. D }{\bf #1},~#3~(#2)}
\def\npb#1#2#3{{Nucl. Phys. }{\bf B#1},~#3~(#2)}
\def\np#1#2#3{{Nucl. Phys. B }{\bf #1},~#3~(#2)}
\def\npps#1#2#3{{Nucl. Phys. B (Proc. Sup.)}
{\bf #1},~#3~(#2)}
\def\mpl#1#2#3{{Mod. Phys. Lett.}
{\bf #1},~#3~(#2)}
\def\arnps#1#2#3{{Annu. Rev. Nucl. Part. Sci.}
{\bf #1},~#3~(#2)}
\def\sjnp#1#2#3{{Sov. J. Nucl. Phys.}
{\bf #1},~#3~(#2)}
\def\jetp#1#2#3{{JETP Lett. }{\bf #1},~#3~(#2)}
\def\app#1#2#3{{Acta Phys. Polon.}
{\bf #1},~#3~(#2)}
\def\rnc#1#2#3{{Riv. Nuovo Cim.}
{\bf #1},~#3~(#2)}
\def\ap#1#2#3{{Ann. Phys. }{\bf #1},~#3~(#2)}
\def\ptp#1#2#3{{Prog. Theor. Phys.}
{\bf #1},~#3~(#2)}
\def\apjl#1#2#3{{Astrophys. J. Lett.}
{\bf #1},~#3~(#2)}
\def\apjs#1#2#3{{Astrophys. J. Suppl.}
{\bf #1},~#3~(#2)}
\def\n#1#2#3{{Nature }{\bf #1},~#3~(#2)}
\def\apj#1#2#3{{Astrophys. J.}
{\bf #1},~#3~(#2)}
\def\anj#1#2#3{{Astron. J. }{\bf #1},~#3~(#2)}
\def\mnras#1#2#3{{MNRAS }{\bf #1},~#3~(#2)}
\def\grg#1#2#3{{Gen. Rel. Grav.}
{\bf #1},~#3~(#2)}
\def\s#1#2#3{{Science }{\bf #1},~#3~(#2)}
\def\baas#1#2#3{{Bull. Am. Astron. Soc.}
{\bf #1},~#3~(#2)}
\def\ibid#1#2#3{{\it ibid. }{\bf #1},~#3~(#2)}
\def\cpc#1#2#3{{Comput. Phys. Commun.}
{\bf #1},~#3~(#2)}
\def\astp#1#2#3{{Astropart. Phys.}
{\bf #1},~#3~(#2)}
\def\epjc#1#2#3{{Eur. Phys. J. C}
{\bf #1},~#3~(#2)}
\def\nima#1#2#3{{Nucl. Instrum. Meth. A}
{\bf #1},~#3~(#2)}
\def\jhep#1#2#3{{J. High Energy Phys.}
{\bf #1},~#3~(#2)}
\def\jcap#1#2#3{{J. Cosmol. Astropart. Phys.}
{\bf #1},~#3~(#2)}
\def\lnp#1#2#3{{Lect. Notes Phys.}
{\bf #1},~#3~(#2)}
\def\jpcs#1#2#3{{J. Phys. Conf. Ser.}
{\bf #1},~#3~(#2)}
\def\aap#1#2#3{{Astron. Astrophys.}
{\bf #1},~#3~(#2)}
\def\mpla#1#2#3{{Mod. Phys. Lett. A}
{\bf #1},~#3~(#2)}

\end{document}